\documentclass[conference]{IEEEtran}
\IEEEoverridecommandlockouts
\usepackage{cite}
\usepackage{amsmath,amssymb,amsfonts}
\usepackage{algorithmic}
\usepackage{graphicx}
\usepackage{textcomp}
\usepackage{xcolor}

\usepackage{soul}
\usepackage{times}
\usepackage{empheq}
\usepackage{latexsym}
\usepackage{booktabs}
\usepackage{multirow}
\usepackage{subfigure}
\usepackage{tabularx}
\usepackage{color}
\usepackage{graphicx}
\usepackage{caption}

\usepackage{float}
\usepackage{rotating}
\usepackage{hyperref}

\def\BibTeX{{\rm B\kern-.05em{\sc i\kern-.025em b}\kern-.08em
    T\kern-.1667em\lower.7ex\hbox{E}\kern-.125emX}}
\begin{document}

\title{Disentangling Specificity for Abstractive Multi-document Summarization\\
}

\author{\IEEEauthorblockN{Congbo Ma\textsuperscript{1}, Wei Emma Zhang\textsuperscript{2}, Hu Wang\textsuperscript{2}, Haojie Zhuang\textsuperscript{2}, Mingyu Guo\textsuperscript{2} }
\IEEEauthorblockA{\textit{\textsuperscript{1}Macquarie University, Australia}  \textit{\textsuperscript{2}The University of Adelaide, Australia}\\
congbo.ma@mq.edu.au,
\{wei.e.zhang, hu.wang, haojie.zhuang, mingyu.guo\}@adelaide.edu.au}
}
\maketitle

\begin{abstract}
Multi-document summarization (MDS) generates a summary from a document set. Each document in a set describes topic-relevant concepts, while per document also has its unique contents. However, the document specificity receives little attention from existing MDS approaches. Neglecting specific information for each document limits the comprehensiveness of the generated summaries. To solve this problem, in this paper, we propose to disentangle the specific content from documents in one document set. The document-specific representations, which are encouraged to be distant from each other via a proposed orthogonal constraint, are learned by the specific representation learner. We provide extensive analysis and have interesting findings that specific information and document set representations contribute distinctive strengths and their combination yields a more comprehensive solution for the MDS. Also, we find that the common (i.e. shared) information could not contribute much to the overall performance under the MDS settings. Implemetation codes are available at \href{https://github.com/congboma/DisentangleSum}{https://github.com/congboma/DisentangleSum}.
\end{abstract}


\section{Introduction}

Multi-document summarization (MDS) is an important task in the natural language processing \cite{ma2022multi}. Processing the source documents by flat-concatenating them into a mega document is one way to solve MDS tasks \cite{Alexander2019multinews, mao2020multi, Chao2022Read}. But the purpose of MDS is to treat each document as a standalone unit, digging out the connections and differences between documents and generating informative and comprehensive summaries. Targeting these issues, some researchers tried to establish the connections not only at word-level relations but also sentence, paragraph and document levels. They employ hierarchical Transformer structures \cite{liu2019hierarchical, li2020leveraging, jin2020multi, Yun2022improving} to forge connections among documents. The high-level Transformer encodes the paragraph representations from different documents. Besides, some existing works incorporated graph information \cite{li2020leveraging, pasunuru2021efficiently, wang2020heterogeneous} to build connections among documents. However, these methods are not specifically designed for extracting specific features and therefore they ignore the specific information contained in each document in a document set.

Nonetheless, the extraction of specific information is crucial with the following reasons: (1) In a collection of documents, each document contains not only the common information but also has specific contents that distinguish it from other documents. These specific information contain unique facts, viewpoints, and details \cite{Alexander2019multinews}. Extracting these specific details enhance the comprehensiveness of the resulting summary. Additionally, some essential information may be exclusive to a particular document, yet it plays a pivotal role in obtaining a comprehensive grasp of the entire document set. 
Therefore, a high-quality MDS summary should not only be able to capture document commonality but can also comprehensively consider the specific information from each document, covering various dimensions to meet the user's demand for a comprehensive understanding of the documents \cite{Ruben2022how}. (2) Focusing on the extraction of specific information helps reduce redundancy, rendering the summary more concise and informative.
Clustering-based MDS methods \cite{wan2008multi, pasunuru2021efficiently, Ori2022proposition} can be used to group similar sentences or pieces of information and remove redundancy. After removing the redundant information, the remaining information in each document can be viewed as implicitly specific information. However, the specificity within these remaining information cannot be explicitly guaranteed to be distinctive between documents.

In order to address this issue, our intuition is not only to capture the overall information in a document set but also to distinguish the specificity of each document and learn representations of document specificity which will be considered in the summary generation process.
To this end, we propose DisentangleSum --- a simple yet effective summarization model that disentangles document uniqueness with a set of document-specific representation learners. In order to optimize the learning of specific representations, we further propose an orthogonal constraint to encourage the specific representations obtained from a pair of documents to be distinctive from each other. Based on the constraint, we design an objective function that can transform the quadratic increment of the losses between each of the paired documents into linear to cope with a large number of documents in a set. We summarize our contributions as follows:

\begin{itemize}

\item We present DisentangleSum, an innovative MDS model that is capable of disentangling specific information from each document in a set, leading to more comprehensive summary generation. To the best of our knowledge, we are the first to consider the specific information for deep learning based MDS task.

\item To incentivize the document-specific learner to retain document specificity information, we propose an orthogonal constraint. This constraint encourages the document-specific representation vectors to align vertically with each other, ensuring a semantic separation between them.

\item Experimental results on two MDS datasets demonstrate the effectiveness of DisentangleSum. 
We additionally offer comprehensive analyses from multiple perspectives to investigate the underlying mechanisms of DisentangleSum and circumstances of the proposed model can work.
\end{itemize}
 
\begin{figure*}[t]
\centering
\includegraphics[width=0.65\textwidth]{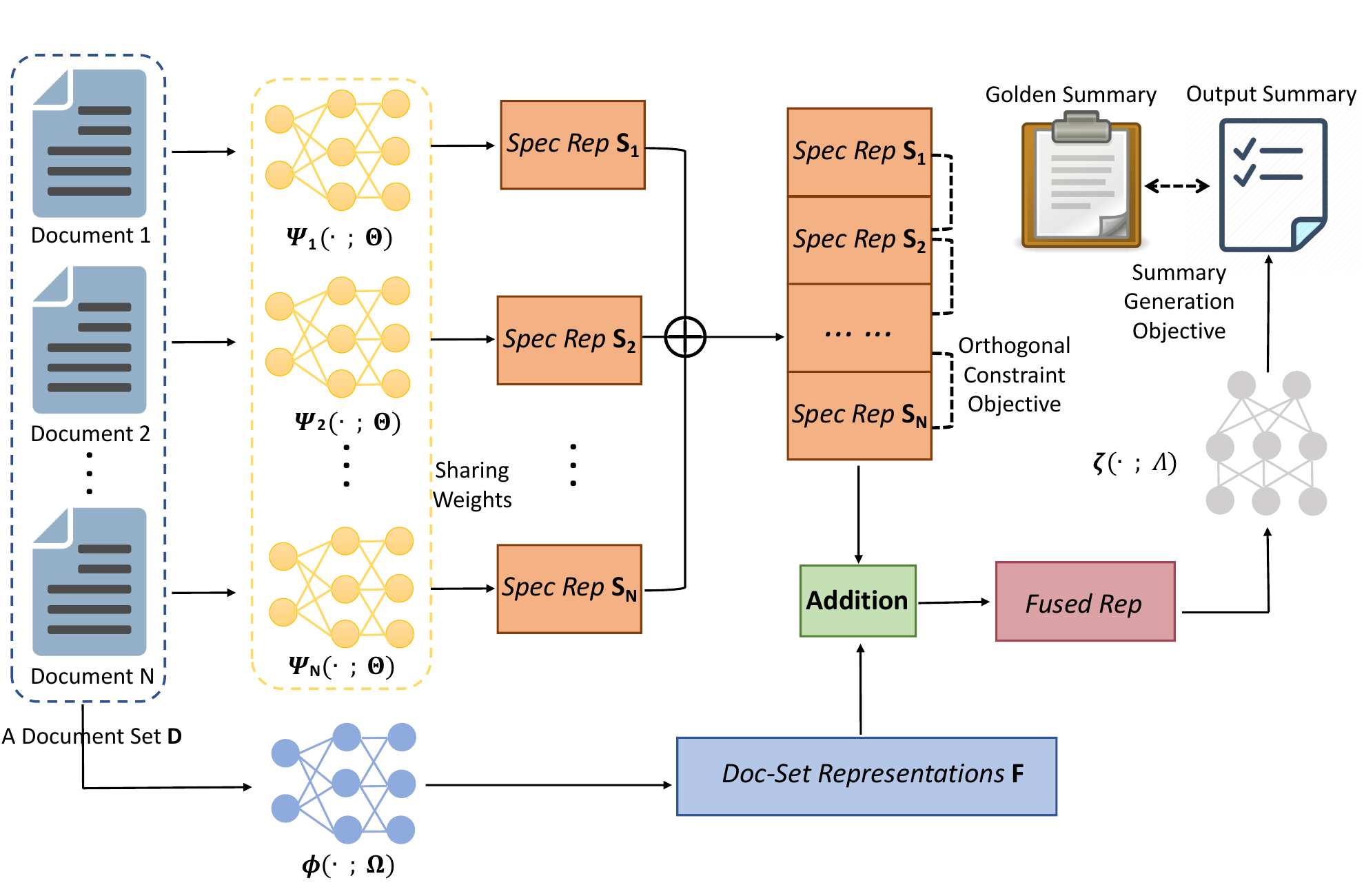}
\vspace{-2mm}
\caption{The overall framework of the proposed DisentangleSum model.}
\label{fig:framework}
\end{figure*}
\vspace{-2mm}

\section{Related Works}

\noindent \textbf{Mining document set representations.} One way to process MDS input is to concatenate documents into a single mega document, utilizing overall document set representations  \cite{Alexander2019multinews, mao2020multi, Chao2022Read}. This concatenation neglected the relative importance of documents, reordering the documents in a document set according to their importance makes it easier for the summarization model to learn \cite{Chao2022Read}. However, Methods deal with mega documents do not explicitly consider interconnections and distinctions among individual input documents \cite{liu2019hierarchical}.

\noindent \textbf{Mining document relation representations.} Researchers have focused on mining relationships among documents in the same set using different techniques. These include extracting graph representations \cite{li2020leveraging, wang2020heterogeneous, zhao2020summpip}, leveraging document-level positional relations \cite{Ma2022document}, and performing entity extraction \cite{Wen2022primera}. These methods incorporate domain-specific knowledge or semantic relationships among the source documents for better performance. Additional approaches to document relation mining include hierarchical Transformer architectures \cite{liu2019hierarchical} and attention mechanisms with different granularity representations \cite{jin2020multi}.
However, the existing methods mentioned above overlook the distinctiveness of individual documents within a document set, inevitably leading to incomprehensive summarization. To fill this research gap, we propose the DisentangleSum model, which extracts document-specific representations for comprehensive summary generation.

\section{Methodology}

In this section, we provide an overview of the proposed model (Figure \ref{fig:framework}), DisentangleSum, by describing how to incorporate document disentangling specificity representation learning into the summarization framework. We propose the orthogonal constraint applied during the training of document-specific representations.

\subsection{Problem Formulation}

In MDS, each document set can have a varying number of documents. For illustration purposes, let's consider a document set $\mathbf{D}=(\mathbf{d}_1, \mathbf{d}_2, \mathbf{d}_3, ..., \mathbf{d}_N)$ consisting of $N$ input documents related to a specific topic or sharing common information. In our approach, we utilize the specific encoder $\psi_i(\cdot;\Theta)$ for document $\mathbf{d}_i$, where $\Theta$ represents the learnable parameters. These specific encoders generate specific representations $S_i$ for each document, and collectively, they form the specific representations $\mathbf{S}$ for the entire document set. Additionally, we employ a document-set encoder $\phi(\cdot;\Omega)$ with learnable parameters $\Omega$ to obtain document-set representations $\mathbf{F}$. The target is to generate a concise summary output $\mathbf{O}$ that synthesizes all important contents from input documents by considering both specific representations $\mathbf{S}$ and document-set representations $\mathbf{F}$.

\subsection{Document Specific Representation Learner}

In a document set, the specific representation learner aims to identify the specific information within each document. To achieve this, we introduce a specific encoder to encode document $\mathbf{d}_i$ in the same document set $\mathbf{D}$:

\vspace{-1mm}
\begin{equation}  \small
\mathbf{S}_i =\psi_i(\mathbf{d}_i;\Theta),
\end{equation}

\noindent Under the setting of MDS, the number of input documents can vary within a document set. 
To address this variability, we propose a design where the learnable parameters, denoted as $\Theta$, are shared across a set of $N$ specific encoders instead of assigning a separate specific encoder to each document in the set. The rationale behind this approach stems from the fact that documents with identical indexes in different document sets are unrelated in content. Consequently, maintaining multiple separate specific encoders for each indexed document is not reasonable.
Subsequently, we concatenate these specific representations to obtain the overall specific representations of a document set:

\vspace{-1mm}
\begin{equation}   \small
\mathbf{S} = \mathbf{S}_1 \oplus \mathbf{S}_2 \oplus \mathbf{S}_3 \oplus ... \oplus \mathbf{S}_N,
\end{equation}

\noindent $\oplus$ is the concatenation operation in this paper. To enable sufficient expressive power for representations to be decoded,
we also obtain the document-set representations $\mathbf{F}$ out of a document set $\mathbf{D}$ by: 

\vspace{-1mm}
\begin{equation}   \small
\mathbf{F}=\phi(\mathbf{D};\Omega) ,
\end{equation}
\noindent Next, we combine the document-set representations and specific representations by performing an element-wise addition and decoding them into summarization outputs:

\vspace{-1mm}
\begin{equation}   \small
\mathbf{O} = \zeta(\alpha \cdot \mathbf{F} + \mathbf{S}; \Lambda),
\end{equation}

\noindent Here, $\alpha$ serves as a trade-off factor to control the weight balancing between the document-set representations and specific representations. The dimension of $\mathbf{F}$ and $\mathbf{S}$ are both equal to the length of a document set.
The decoder function $\zeta(\cdot; \Lambda)$, parameterized by $\Lambda$, is responsible for decoding the intermediate representations into concise summaries.

\subsection{Orthogonal Constraint within the Training of Document Specific Representations}
\label{sec:circle-paired-loss}

To guide specific representations learning, we impose an orthogonal constraint between pairs of specific representations $\mathbf{S}_i$ and $\mathbf{S}_j$. The document-specific loss, which promotes dissimilarity between specific representations, is defined as:

\vspace{-1mm}
\begin{equation}   \small
 L_{\mathit{spec}} = \sum_{i}^{} \sum_{j}^{} \left \| \mathbf{S}_i^{\ \top} \ \mathbf{S}_j \right \| _{F}^{2} ,
 \label{equ:spec_loss}
\end{equation}
\noindent where $\left \| \cdot \right \| _{F}^{2} $ is the squared Frobenius norm. 
To encourage dissimilarity between specific representations, we aim for a smaller inner product between each pair of specific representation vectors, promoting orthogonality. This ensures distinctiveness among specific representations within the same document set. As the specific encoder learns, it captures each document's unique essence, thereby retaining specific content. However, when a document set contains more than two documents, the computation of specific representation objectives between every documents pair grows quadratically. To address this, we introduce a circle-paired loss objective function, reducing complexity from quadratic to linear. Formally, we have:

\vspace{-1mm}
\begin{equation}  \small
\label{equ:spec}
L_{\mathit{spec}}=\sum_{i=1}^{N} L_{\mathit{spec}}^{i} \ ,
\end{equation}

\vspace{-1mm}
\begin{equation}  \small
    L_{\mathit{spec}}^{i}  = 
    \begin{cases}
    \left \| \mathbf{S}_i^{\ \top} \ \mathbf{S}_{i+1} \right \| _{F}^{2} & i\ne N \\
    \left \| \mathbf{S}_N^{\ \top} \ \mathbf{S}_{1} \right \| _{F}^{2} & i = N
    \end{cases}, \label{equ:speci}
\end{equation}

The objective function calculates the specific representation costs between each document and the subsequent document in the set, with the last document computed against the first one.

\subsection{Overall Objectives}

The proposed framework aims to train a high-quality summarization model that incorporates specific representations from each document. This is achieved through two key components: an orthogonal constraint for distinct document representations and a supervised cross-entropy loss concerning golden summaries:

\vspace{-1mm}
\begin{equation}  \small
L_{\mathit{total}} = L_{\mathit{gen}} + \beta \cdot L_{\mathit{spec}},
\end{equation}

\vspace{-1mm}
\begin{equation}   \small
L_{\mathit{gen}} = -\sum_{k=1}^{M} \eta(\hat{\mathbf{O}}_k,\mathbf{O}_k)\log (p(\mathbf{O}_k)),
\end{equation}

\noindent where $\beta$ is a balance factor, $p(\mathbf{O}_k)$ is one shard\footnote{Following the implementation in https://github.com/Alex-Fabbri/Multi-News/blob/master/code/OpenNMT-py-baselines/onmt/utils/loss.py, shards are segments when computing losses.} of the predictive summary from the DisentangleSum model. $\hat{\mathbf{O}}_k$ denotes corresponding true labels. $M$ represents the number of shard within the generated summary. The calculation of $\eta(\cdot,\cdot)$ in a summarization task is different from other tasks such as text classification. $\eta(\cdot,\cdot)$ indicates the evaluation function between prediction and ground truth, the widely used ROUGE evaluation are adopted here.

\section{Experiments}

\subsection{Datasets \& Evaluation Metrics \& Baselines}

We assess the effectiveness of the proposed method on Multi-News\cite{Alexander2019multinews} and Multi-XScience \cite{Yao2020multi} datasets which satisfy that documents contain the specific information in a document set. Both datasets are truncated to 500 tokens. We use standard summarization evaluation metrics ROUGE\footnote{The parameters of ROUGE are -c 95 -2 -1 -U -r 1000 -n 4 -w 1.2 -a -m.} \cite{lin2004rouge} BERTScore\footnote{The model type of BERTScore is bert-base-uncased.} \cite{Zhang2020BERTScore} 
Additionally, we employ the coverage score \cite{Max2018Newroom} to quantify the amount of information retained in the generated summaries compared to input documents.
We compare with the following strong baselines: \textit{LexRank} \cite{gunes2004lexrank}, \textit{TextRank} \cite{rada2004textrank}, \textit{MMR} \cite{Jaime1998mmr}, \textit{BRNN}, \textit{Vanilla Transformer (VT)} \cite{ashish2017attention} and its variant \textit{CopyTransformer (CT)}, \textit{Pointer-Generator (PG)} \cite{seelm2017pg}, \textit{Hi-MAP} \cite{Alexander2019multinews}, \textit{Hierarchical Transformer (HierTrans)} \cite{liu2019hierarchical}, \textit{SummPip} \cite{zhao2020summpip}, \textit{SAGCopy} \cite{xu2020self}, \textit{HeterGraphSum (HGS)} \cite{wang2020heterogeneous}, \textit{Highlight-Transformer (HiTrans)}\cite{liu2021highlight}, \textit{DocLing} \cite{Ma2022document}.

\subsection{Implementation Details}

During model training, the initial learning rate is set to 2. The training strategy involves a warm-up phase for the first 8,000 steps, followed by multi-step learning rate reduction. The batch size is set to 4,096, and the models are trained for 20,000 steps using the \textit{Adam} optimizer. Both the encoder and decoder consist of four transformer layers, and positional encoding is applied. The dropout rate is 0.2. The trade-off factor for specific representations ($\alpha$) is set to 0.01, and the trade-off factor for specific loss ($\beta$) is set to 0.001. The word embedding size for source documents is set to 512 dimensions. 
We conduct all the experiments on one NVIDIA 3090 GPU with one Intel i9-10900X CPU upon Ubuntu 22.04.3 LTS Operation System.
For the minimum and maximum lengths of generated summaries, Multi-News has 200 and 300 words, while Multi-XScience has 110 and 300 words.

\begin{table}[t] \small
\centering
\scalebox{0.95}{
\setlength{\tabcolsep}{3mm}{
\begin{tabular}{l|cc}
\hline
\hline
Models & \begin{tabular}[c]{@{}c@{}}Multi-News\end{tabular} & \begin{tabular}[c]{@{}c@{}}Multi-XScience\end{tabular} \\ \hline
VT     & 19.76 & 18.98 \\
CT & 46.78 & 15.82 \\
DocLing          & 49.62 & 20.10 \\ \hline
DisentangleSum        & \textbf{52.99}& \textbf{22.68} \\    \hline \hline
\end{tabular}}}
\vspace{-1mm}
\caption{Coverage score comparison.}
\label{tab:coverage}
\end{table}

\begin{table}[]  \small
\centering
\scalebox{0.95}{
\setlength{\tabcolsep}{3mm}{
\begin{tabular}{l|cccc}
\hline
\hline
Models        & R-1   & R-2   & R-SU  & BS                             \\ \hline
LexRank       & 37.92 & 13.1  & 12.51 & 0.83                           \\
TextRank      & 39.02 & 14.54 & 13.08 & 0.83                           \\
MMR           & 42.12 & 13.19 & 15.63 & 0.84                           \\
BRNN          & 38.36 & 13.55 & 14.65 & 0.83                           \\
VT  & 25.82 & 5.84  & 6.91  & 0.80                            \\
CT     & 42.98 & 14.48 & 16.91 & 0.84                           \\
PG            & 34.13 & 11.01 & 11.58 & 0.83                           \\
Hi-MAP        & 42.98 & 14.85 & 16.93 & 0.83                           \\
HierTrans     & 36.09 & 12.64 & 12.55 & 0.84                           \\
SummPip       & 42.29 & 13.29 & 16.16 & 0.84                           \\
SAGCopy       & 43.98 & 15.21 & 17.65 & -                              \\
HGS & 43.62 & 14.99 & 17.29 & \textbf{0.85} \\
HiTrans       & 44.62 & 15.57 & 18.06 & -                              \\
DocLing       & 44.35 & 15.04 & 17.97 & \textbf{0.85} \\ 
\hline
DisentangleSum & \textbf{45.95} & \textbf{16.32} & \textbf{19.23} & \textbf{0.85} \\ 
\hline
\hline
\end{tabular}}}
\vspace{-1mm}
\caption{Performance comparison on the Multi-News.} 
\label{tab:overall_performance_Multi-News}
\end{table}

\begin{table} [t] \small
\centering
\scalebox{0.95}{
\setlength{\tabcolsep}{3mm}{
\begin{tabular}{l|cccc}
\hline
\hline
Models       & R-1   & R-2                            & R-SU & BS                             \\ \hline
LexRank      & 31.31 & 5.85                           & 9.13 & 0.83                           \\
TextRank     & 31.15 & 5.71                           & 9.07 & \textbf{0.84} \\
MMR          & 30.04 & 4.46                           & 8.15 & 0.83                           \\
BRNN         & 27.95 & 5.78                           & 8.43 & 0.83                           \\
VT & 28.34 & 4.99                           & 8.21 & 0.82                           \\
CT    & 26.92 & 4.92                           & 7.50  & 0.83                           \\
PG           & 30.30  & 5.02                           & 9.04 & \textbf{0.84} \\
Hi-MAP       & 30.41 & 5.85                           & 9.13 & 0.81                           \\
HierTrans    & 25.31 & 4.23                           & 6.64 & 0.83                           \\
SummPip      & 29.66 & 5.54                           & 8.11 & 0.82                           \\
DocLing      & 30.93 & \textbf{6.06} & 9.57 & 0.84                           \\ \hline
DisentangleSum & \textbf{31.81} & 5.90 & \textbf{9.88} & \textbf{0.84} \\ \hline
\end{tabular}}}
\vspace{-1mm}
\caption{Performance comparison on the Multi-XScience.}
\label{tab:overall_performance_MultiXScience}
\end{table}

\begin{table}[t] \small
\centering
\scalebox{0.95}{
\setlength{\tabcolsep}{3mm}{
\begin{tabular}{@{}l|cccc@{}}
\hline
\hline
Models      &   Speci & Compr & Coher & Relev \\ \hline
VT   & 1.67 &2.28  &2.13 & 2.46    \\
CT   &2.33 &2.70  &2.27  &2.78     \\
DocLing       & 2.89  &3.10  &2.78  &2.93      \\
\hline
DisentangleSum    & \textbf{3.16} &\textbf{3.87} &\textbf{3.21} &\textbf{3.13 }           \\
\hline
\hline
\end{tabular}}}
\vspace{-1mm}
\caption{Human evaluation results on the Multi-News dataset. }
\label{tab: human-evaluation}
\end{table}

\subsection{Main Results} 
This section is designated for validating the model's effectiveness from three perspectives: (1) verifying the comprehensiveness of the generated summaries; (2) evaluating the overall performance through automated evaluations; (3) evaluating human feedback on specific information extraction, comprehensive, coherence and relevance.

\begin{table}[h] \LARGE
\centering
\scalebox{0.46}{
\begin{tabularx}{\textwidth}{|l|X|}
\hline 
Doc Set &  \textbf{\textcolor{red} {Doc \#1}}: \textcolor{blue}{the world's ugliest color has been described as "death," "dirty" and "tar," but this odious hue is serving an important purpose: discouraging smoking.} ... the agency was hired by the australian government to find a color that was so repugnant that if it was on tobacco products, it would dissuade ... \newline 
\textbf{\textcolor{red} {Doc \#2}}: ... changes to australia ’ s duty free tobacco allowance smoking prevalence rates abs national aboriginal and torres strait islander social survey, \textcolor{green}{2014-15 the proportion of aboriginal and torres strait islander people aged 15 years and over who were daily smokers was 38.9 \% in 2014-15 , down from 44.6 \% in 2008 and 48.6 \% in 2002} ... \newline
\textbf{\textcolor{red} {Doc \#5}}:  ... \textcolor{orange}{in may, previously passed legislation will go into effect requiring all packs of cigarettes to be standardized. tobaccos products will be stripped of brightly colored branding and replaced with a sludge-like color .} . but does the stripped-down , " ugly " packaging really reduce smoking ...
\\ 
\hline
VT  & –australia's news agency says it's time to get rid of certain types of \textlangle unk\textrangle products. the australian government has approved a ban on <unk> products , which include ...\\
\hline
CT & -\textcolor{blue} {the world's ugliest color will be helpful in smoking rates in their country, according to a team of experts} ... researchers found that pantone publications, including \textcolor{blue}{pantone 448c visually, are chock full of " ugly " reactions}, including ...\\
\hline
\begin{tabular}[c]{@{}l@{}}DocLing\\ \end{tabular}  & ... \textcolor{blue}{world's ugliest color} - \textcolor{orange} {will be stripped of colored branding and replaced with a sludge-like color .in may, previously passed legislation will go into effect requiring all packs of cigarettes to be standardized} ... \\
\hline
\begin{tabular}[c]{@{}l@{}}Ours\end{tabular} & \textcolor{blue} {–the world's ugliest color is serving an important purpose}: ... \textcolor{blue} {more likely to deter smoking from reaching for their next pack of cigarettes} ... \textcolor{orange}{it will go into effect requiring to be standardized} ... \textcolor{green}{found that smokers aged 15 years and over half a year} ... \textcolor{green}{in smoking rates than those in 2002 and 48.6 \% in 2002} ... \\
\hline
\end{tabularx}}
\caption{Example of source documents and summaries.}
\label{tab:summary-examples}
\end{table}


\vspace{1mm}
\subsubsection{Coverage Score}
We conduct a comparison based on coverage score (Table \ref{tab:coverage}) among three Transformer-based models: Vanilla Transformer, CopyTransformer, and DocLing. Despite sharing a similar structure, these models lack consideration for document specificity. 
Coverage score measures the percentage of words in the generated summary that come directly from the source documents. A higher score indicates that the model can generate summaries with richer information from the source documents.
From the results, disentangleSum outperforms counterparts by a large margin, achieving the highest coverage score on both datasets. This indicates its ability to generate more comprehensive summaries, preserving substantial information from the original documents.

\vspace{1mm}
\subsubsection{Overall Performance}

Table \ref{tab:overall_performance_Multi-News} and Table \ref{tab:overall_performance_MultiXScience}\footnote{ The code for SAGCopy \cite{xu2020self} and HiTrans \cite{liu2021highlight} is not publicly available and they did not provide results on the Multi-XScience dataset. Additionally, since MultiX-Science does not contain labels for extractive summarization, hindering the the HeterGraphSum \cite{wang2020heterogeneous} from being implemented on it.} shows the DisentangleSum model receives outstanding performance in most of the cases. 
On the Multi-News dataset, DisentangleSum outperforms the second-best model, attaining 1.6 improvement on ROUGE-1, 1.28 improvement on ROUGE-2, and 1.26 improvement on ROUGE-SU. Particularly, the ROUGE-SU score received 7\% improvement over the second-best model.
Similarly results are shown on Multi-XScience data as well. 
The superior results can be consistently gained because the proposed DisentangleSum model has been enhanced with the capability to capture both the document set and document-specific features, leading to a better summary generation.

\vspace{1mm}
\subsubsection{Human Evaluation}

We conduct human evaluations to assess summary quality in terms of Specificity (Speci), Comprehensiveness (Compr), Coherence (Coher), and Relevance (Relev), aiming to detect diverse viewpoints from multiple documents.
Three Ph.D students in NLP examine the performance of four models using 50 randomly sampled source documents from the Multi-News dataset, scoring from one to five (one = very bad; five = very good). 
The final scores (shown in Table \ref{tab: human-evaluation}), averaged across different cases and raters, consistently favor DisentangleSum across all four metrics.
An example (In Table \ref{tab:summary-examples}) from the MultiNews dataset \cite{Alexander2019multinews} discusses how to discourage smoking, with each document covering the topic from different perspectives.
Doc \#1  indicates the ugliest colour serves an important purpose, Doc \#2 lists the statistics related to smoking, and Doc \#5  discusses smoking from a legal point of view. Existing works provide summaries that miss some specific information. For example, the summary generated by the DocLing \cite{Ma2022document} model fails to include the statistical information presented, resulting in the omission of important specific details from the source documents. These indicate the summaries generated by our model can cover more specific information from the source documents, and exhibit better coherence and relevance.

\begin{table}[t] \small
\centering
\scalebox{0.97}{
\setlength{\tabcolsep}{3mm}{
\begin{tabular}{@{}l|ccc@{}}
\hline
\hline
Objectives & R-1   & R-2   & R-SU  \\ \hline
SSL     &44.64  & 15.47  & 17.83  \\
TL    &44.15  & 14.98  & 17.52  \\
\hline
SL     &44.10 & 15.00 & 17.47  \\
CPL    & 45.16 & 15.39 & 18.48 \\
CPL-R  & 45.03  & 15.09  & 18.40  \\
DPL     & 44.74 & 15.33 & 18.05 \\
DPL-N     &  44.41 &14.93 & 17.86 \\
\hline
\hline
\end{tabular}}}
\vspace{-1mm}
\caption{Models performance with different objective functions on Multi-News validation dataset.``R'' and ``N'' indicates randomly sort documents in the same document set and normalization.}
\label{tab:share_circle_overall}
\end{table}

\subsection{Objective Function Selections}
\label{sec:densecircle}

We evaluate model effectiveness by integrating shared and specific representations using various objective functions. This evaluation aims to highlight capturing common and specific information in MDS and understand how different training objectives affect model performance.

\begin{figure*}[htbp]
\centering
\subfigure{
\begin{minipage}[t]{0.42\linewidth}
\centering
\includegraphics[width=1\textwidth]{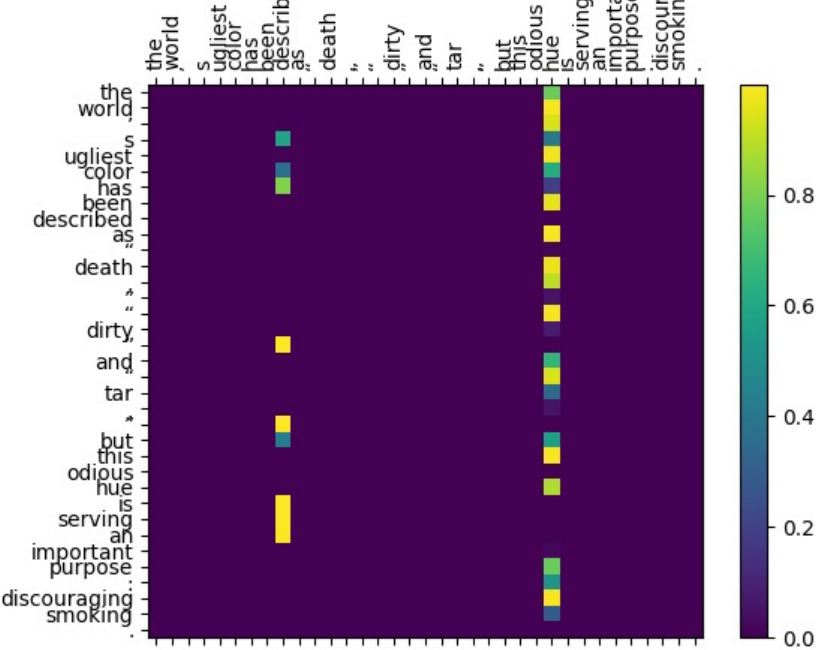}
specific representations
\end{minipage}%
}%
\subfigure{
\begin{minipage}[t]{0.42\linewidth}
\centering
\includegraphics[width=1\textwidth]{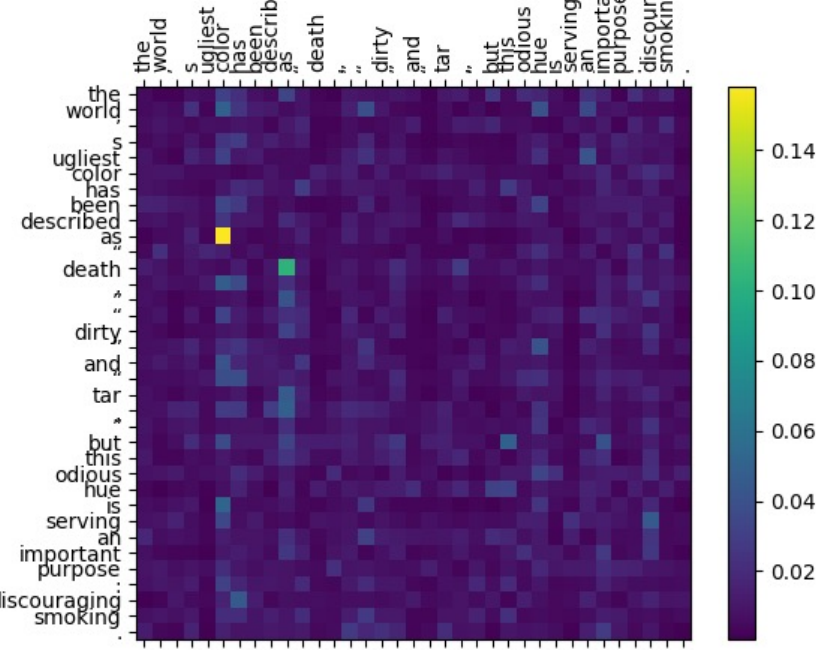}
shared representations
\end{minipage}%
}%
\centering
\vspace{-2mm}
\caption{Attention maps of learned specific representations and shared representations. }
\label{fig:heatmap}
\end{figure*}

\vspace{1mm}
\subsubsection{Specific-Shared Loss ((SSL)) V.S. Triplet Loss (TL)}
Inspired by the good performance of specific representations, we intuitively think that disentangling the shared representation may further improve the model performance. Here shared representation refers to common information shared by multiple documents in one set. 
To obtain the shared representations, we incorporate a shared encoder $\omega(\cdot;\nu)$ to learn the shared representations $\mathbf{H_{i}}$ from the document $\mathbf{d_{i}}$ by: 

\vspace{-1mm}
\begin{equation}   \small
\mathbf{H_{i}}=\omega(\mathbf{d_{i}};\nu ),
\end{equation}
\noindent where $\nu$ is a set of learnable parameters. Within a set, shared representations should be similar, while specific representations should be distinguishable. We use two objective functions to achieve this: one to encourage similarity in shared representations and another to emphasize distinctiveness in specific representations.
Specific-shared loss represents by:

\vspace{-1mm}
\begin{equation}   \small 
L_{\mathit{total}} = L_{\mathit{gen}} + \beta \cdot L_{\mathit{spec}}+\gamma \cdot L_{\mathit{shared}} \ ,
\end{equation}

\vspace{-1mm}
\begin{equation}  \small
\label{equ:shared}
L_{\mathit{shared}}=\sum_{i=1}^{N} L_{\mathit{shared}}^{i} \ ,
\end{equation}

\vspace{-1mm}
\begin{equation}  \small
   L_{\mathit{shared}}^{i}  = 
    \begin{cases}
    \left \| \mathbf{H}_i -\mathbf{H}_{i+1} \right \| _{p} & i\ne N \\
    \left \| \mathbf{H}_N - \ \mathbf{H}_{1} \right \| _{p} & i = N
    \end{cases},
\end{equation}

\vspace{-1mm}
\begin{equation}   \small
 L_{\mathit{spec}} = \sum_{i=1}^{N} \left \| \mathbf{S}_i^{\ \top} \ \mathbf{H}_i \right \| _{F}^{2},
\end{equation}

\noindent where $\gamma$ is a balance factor. For a document $d_{i}$, we expect its specific representations to be orthogonal with its shared representations.

Triplet loss is inspired by contrastive representation learning from positive and negative samples \cite{Florian2015FaceNet}. For a document $d_{i}$, its shared and specific representation,  $\mathbf{H_{i}}$ and  $\mathbf{S_{i}}$, can be seen as an anchor and a negative sample. The positive sample can be another shared representation map $\mathbf{H_{j}}$ from document $d_{j}$. Note that $d_{i}$ and $d_{j}$ are in the same document set. The final objective function can be:

\vspace{-1mm}
\begin{equation}   \small
L_{\mathit{total}} = L_{\mathit{gen}} + \beta \cdot L_{\mathit{triplet}} \ ,
\end{equation}

\vspace{-1mm}
\begin{equation}  \small
L_{\mathit{triplet}}=\sum_{i=1, j\ne i}^{N}  max (\left \| \mathbf{H}_i -\ \mathbf{H}_j \right \| _{2}^{2}+\left \| \mathbf{H}_i- \ \mathbf{S}_i \right \| _{2}^{2}) \ ,
\end{equation}

Table \ref{tab:share_circle_overall} reveals the performance of the SSL-equipped model surpasses that of the TL-equipped one but  does not surpass the results achieved by the Disentangle model's objective. We hypothesize the conflict between shared and document-set representations causes this phenomenon during the optimization of summary generation. As a result, we decide not to incorporate the shared and specific representations together.

\begin{figure}[htbp]
\centering
\includegraphics[width=0.42\textwidth]{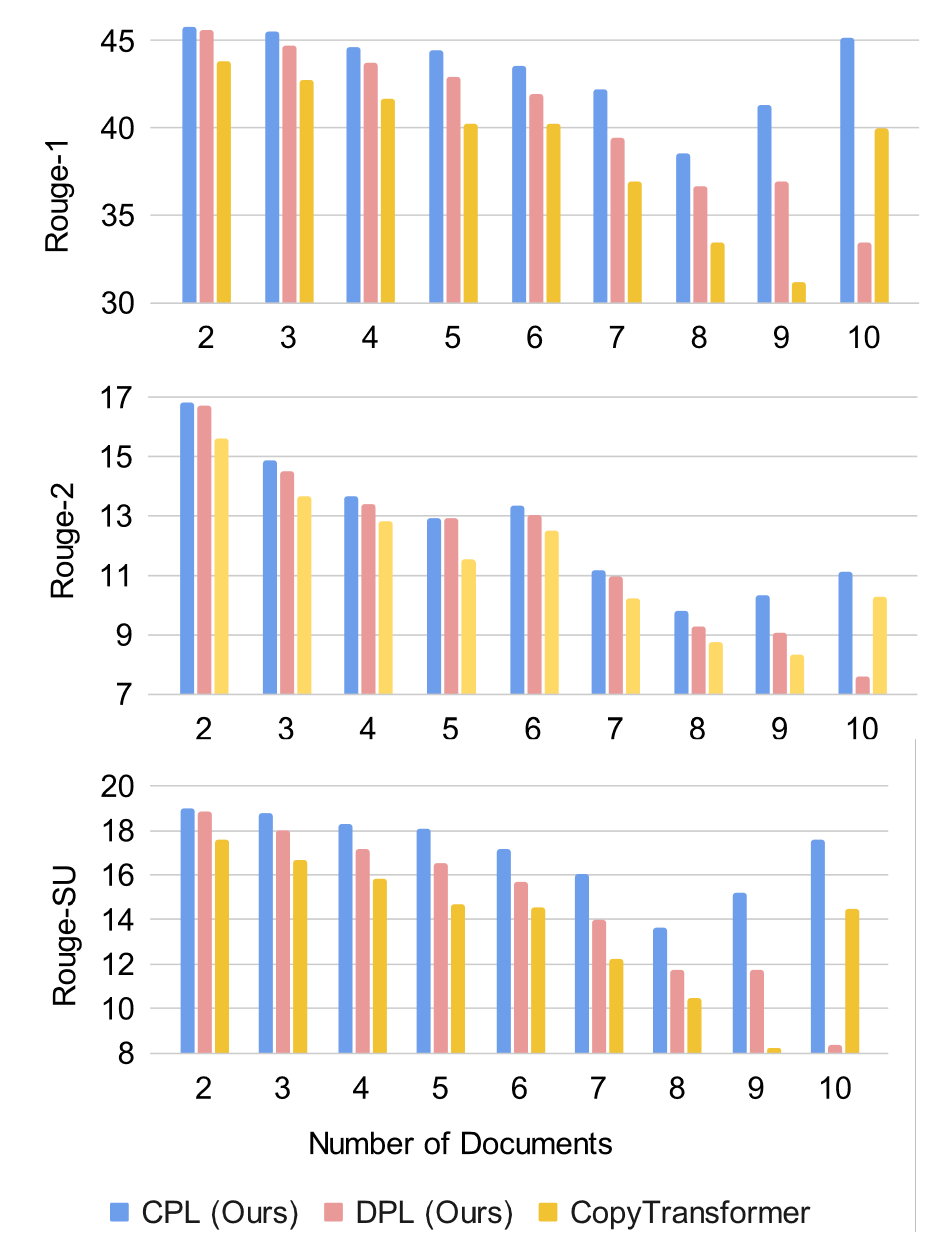}
\vspace{-1mm}
\caption{ROUGE scores of DisentangleSum with circle-paired-loss (CPL) and dense-paired-loss (DPL), and CopyTransformer on document sets containing two to ten documents.}
\label{fig:densecircleloss_R1}
\end{figure}

\begin{table*}[]  
\centering
\scalebox{1}{
\setlength{\tabcolsep}{3mm}{
\begin{tabular}{@{}l|ccc||l|ccc@{}}
\hline
\hline
Variants      & R-1   & R-2   & R-SU  & Variants      & R-1   & R-2   & R-SU  \\ \hline
$\alpha$ = 1        & 43.60  & 13.94 & 17.30  &$\beta$ = 1        & 44.15 & 15.17 & 17.72\\
$\alpha$ = 0.1      & 43.36 & 13.67 & 17.26 &$\beta$ = 0.1      & 44.24 & 15.16 & 17.77 \\
$\alpha$ = 0.01     & 45.16 & 15.39 & 18.48  &$\beta$ = 0.01     & 44.64 & 15.45 & 18.09\\
$\alpha$ = 0.001      & 44.67 & 15.50  & 17.90 &$\beta$ = 0.001      & 45.16 & 15.39 & 18.48\\
w/o Spec Feat & 43.66 & 14.79 & 17.39  &w/o Spec Loss & 43.66 & 14.79 & 17.39 \\ \hline
\hline
\end{tabular}}}
\vspace{-1mm}
\caption{Model performance on Multi-News validation set by tuning specific representation trade-off factor $\alpha$ and loss trade-off factor $\beta$.}
\label{tab:SF}
\end{table*}

\vspace{1mm}
\subsubsection{Specific Loss V.S. Shared Loss (SL)}
Given the better performance of specific-shared loss than triplet loss, we further examine the specific representations and the shared representations separately. We compare the results of specific loss and shared loss. The total objective function to generate shared representations can be defined as:

\vspace{-1mm}
\begin{equation}   \small
L_{\mathit{total}} = L_{\mathit{gen}} + \gamma \cdot L_{\mathit{shared}} \ ,
\end{equation}

\noindent where the calculation of $L_{\mathit{shared}}$ is equal to Equation (\ref{equ:shared}).
In Table \ref{tab:share_circle_overall} shows that the model equipped with shared loss obtains lower performance than that equipped with specific loss.
Furthermore, to dig out why the specific loss has a comparative advantage in summaries generation, we visualize the attention maps (Figure \ref{fig:heatmap}) of specific and shared representations from the last encoding layer. Interestingly, the attention-specific encoder is mainly focused on the individual words ``hue" which is the specific information for \#1 document. However, the heatmap of shared representations is more scattered than the specific representations. This may be because specific representations concentrate on important information of each document while shared representations do not. Consequently, we opt not to select the objective function associated with shared representations for our main experiment.

\vspace{1mm}
\subsubsection{The Selection of Specific Loss}
\label{sec:specloss_selection}

This section investigates two design options for specific loss:
circle-paired loss (CPL), introduced in section \ref{sec:circle-paired-loss}, and dense-paired loss (DPL). DPL involves computing specific representation loss for each pair of documents in the same document set. We conduct two experiments in this subsection:

\noindent  \textbf{(1) Compare the overall performance of DisentangleSum equipped with CPL and DPL. }
We evaluate the models on the Multi-News dataset, analyzing their performance with different objectives. The results in Table \ref{tab:share_circle_overall} indicate that CPL outperforms DPL across all three evaluation metrics.

\begin{figure*}[]
\centering
\subfigure{
\begin{minipage}[t]{0.35\linewidth}
\centering
\includegraphics[width=1\textwidth]{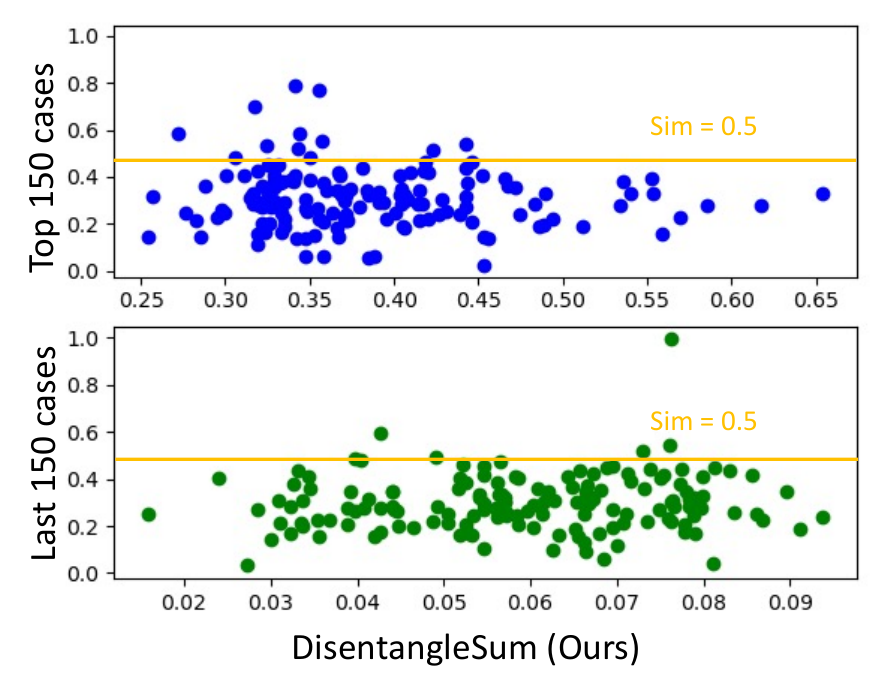}
\end{minipage}%
}%
\subfigure{
\begin{minipage}[t]{0.35\linewidth}
\centering
\includegraphics[width=1\textwidth]{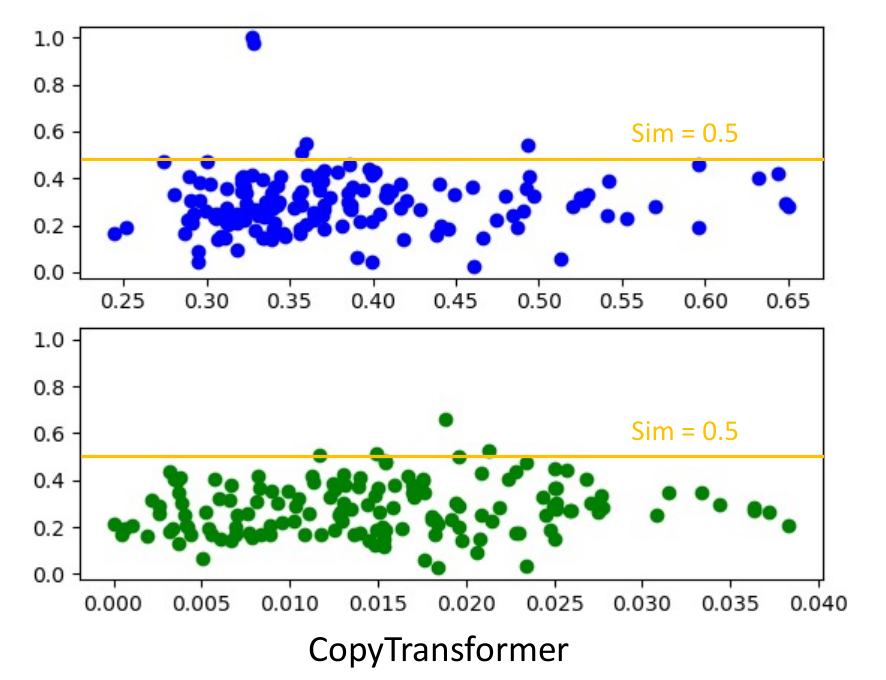}
\end{minipage}%
}%
\centering
\vspace{-2mm}
\caption{The distribution of document similarity scores in the Top 150 and Last 150 cases. The X-axis and Y-axis of each sub-figure are ROUGE-SU scores (scale to 0 $\sim$ 1) and documents similarity scores, respectively. The orange line represents the document similarity score equal to 0.5. }
\label{fig:hardeasy_example}
\end{figure*}

\noindent  \textbf{ (2) Explore the impact of document number on specific loss.} To investigate the relationship between specific loss and the number of documents, we divided the Multi-News validation set into subsets based on the document set size. We compare the model performance trained with CPL and DPL on these subsets.
Figure \ref{fig:densecircleloss_R1} illustrates that DisentangleSum with CPL outperforms DPL and CopyTransformer across all subsets in terms of three ROUGE scores. Notably, when the document set size is two, the results of DPL and CPL are quite similar for the ROUGE-1 score. However, as the number of documents increases, the model trained with DPL experiences a significant performance drop, while the model trained with CPL exhibits a slower decline. This trend holds for ROUGE-2 and ROUGE-SU scores as well. Besides, from the perspective of computational complexity, as the number of documents increases, the document pairs in DPL increase quadratically (e.g. 10 documents yield 45 pairs), while CPL does not.

In order to exclude the impacts of the order of documents in a document set and the loss scale, we further conduct two experiments: (1) Based on CPL, we randomly sort documents in the same document set; (2) Based on DPL, we adjust the loss scale through normalization, dividing the right head side of Equation \ref{equ:spec_loss} by $N^2$. The results (in Table \ref{tab:share_circle_overall}) ruled out the interference of these two factors. The performance differences may be that for MDS tasks, the documents in the same document set describe topic-relevant concepts, yet with some document-specific information. The constraint may be too strong by imposing a model to learn document-specific representations completely different between documents, which in turn may incur a confused model and less “informative” representations learned.

\subsection{Hyperparameter Scale of Models}
We perform a hyperparameter study to examine the effectiveness of specific representation trade-off factor $\alpha$ and loss trade-off factor $\beta$, controlling the trade-off strength of fetching the document specific information and document-set information. The results are shown in Table \ref{tab:SF}. Both the weights of $\alpha$ and $\beta$ are controlled by searching the grid [1, 0.1, 0.01, 0.001, 0]. The experiments of evaluation $\alpha$ is performed under $\beta$ equals to 0.001; while the examination $\beta$ is conducted by setting $\alpha$ to 0.01.
By setting either specific representation weights or specific loss weights to 0s, the model performance is significantly degraded. It suggests the positive contribution of grasping documentary unique information. 
Interestingly, with the increasing of specific representation trade-off factor $\alpha$ from 0.001 to 0.01, the ROUGE scores generally have an increasing trend. But the score goes up and then goes down when $\alpha$ is from 0.01 to 1. The optimal choice of the hyper-parameter $\alpha$ falls in the middle of the evaluated values, which is 0.01. Similar results show for the experiments of loss trade-off factor $\beta$. Generally, 0.001 is recommended for $\beta$ to achieve the best performance. The experimental results indicate that the existence of the document-specific representation learner and the orthogonal constraint of document-specific representation generation is important. Meanwhile, setting large $\alpha$ and $\beta$ obstructs model optimization and summary generation.

\subsection{DisentangleSum Performances with Different Inter-document Similarities}

The aim of this subsection is to analyze the correlation between DisentangleSum performance and inter-document similarities. We define a simple function to calculate the document similarity within a document set using statistical analysis: 

\vspace{-1mm}
\begin{equation}   \small
Sim(\mathbf{D})=\sum_{i=1}^{N-1} \sum_{j=i+1}^{N}\frac{2 \cdot overlap(\mathbf{d}_i, \mathbf{d}_j)}{N(N-1)},
\end{equation}

\noindent where $N$ represents the document number in each document set. It calculates the content overlap between each pair of documents within the document set.
We evaluate the DisentangleSum and CopyTransformer models by calculating ROUGE scores for document sets and ranking them.  We analyze the Top 150 and Last 150 cases, finding average similarity scores of 0.308 and 0.299 for DisentangleSum, and 0.293 and 0.281 for CopyTransformer. 
Figure \ref{fig:hardeasy_example} shows:
(1) DisentangleSum's Top 150 cases have slightly higher document similarity scores than the Last 150 cases. (2) DisentangleSum's Top 150 cases have more instances with similarity scores above 0.5 compared to the Last 150 cases and CopyTransformer's Top 150 cases.

These findings suggest that the proposed model tends to perform better when the document similarity score is higher. The potential reason is in one document set when the overlap between each document is relatively large, the ratio of the uniqueness of each document is relatively small. Models that do not explicitly capture document-specific information may struggle to capture the specific details from the source documents. The DisentangleSum model, designing to retain document-specific information, performs better in such cases when the document similarity score is higher.

\section{Conclusion}

In this paper, we introduce DisentangleSum, a framework to disentangle document-specificity for better abstractive multi-document summarization representations. To optimize the specific representation learning, we apply an orthogonal constraint to encourage the document-specific representation learner to catch specific information per document. The experiments on two prevalent datasets show the superior performances of the proposed model over other counterparts. Furthermore, we also provide extensive analyses that reveal DisentangleSum exhibits broader coverage of input documents and better preservation of document-related information.
These analyses help researchers understand the intuitiveness of the proposed model and could serve as an informative reference to the multi-document summarization research community.

\section*{Acknowledgments}
This research is partially supported by Australian Research Council (ARC) Discovery Project DP230100233.

\bibliography{IEEEexample}
\bibliographystyle{IEEEtran}

\end{document}